\input{aipcheck}

\documentclass[
    ,final            
  ]
  {aipproc}

\layoutstyle{6x9}

\begin{document}

\title{Quasi-elastic Neutrino Scattering -- an Overview}


\author{Jan T. Sobczyk}{
  address={Institute of Theoretical Physics\\ Wroc\l aw University\\ Plac 
Maksa Borna 9, 50-204 Wroc\l aw, Poland}
}

\begin{abstract}
A non-technical overview of charge current quasi-elastic neutrino interaction is presented. 
Many body computations of multinucleon ejection which is proposed 
to explain recent large axial mass measurements are discussed. A few comments 
on recent experimental results reported at NuInt11 workshop are included.
\end{abstract}

\maketitle


\section{Introduction}

Charge current quasi-elastic (CCQE) scattering is the most abundant neutrino 
interaction in experiments like MiniBooNE (MB) or T2K with a flux
spectrum peaked below $1$~GeV. Its full understanding
is crucial for detail neutrino oscillation measurements. 

The very definition of what does the term {\it CCQE} mean 
requires clarifications.
In the case of neutrino-free nucleon scattering the reaction is:

\begin{equation}
\nu + n \rightarrow l^- + p\qquad {\rm or}\qquad 
\bar \nu + p \rightarrow l^+ + n\label{definition}
\end{equation}
with $\nu$, $\bar\nu$, $l^\pm$, $p$ and $n$ standing for: neutrino, antineutrino, charged lepton, proton and neutron respectively.
In the case of neutrino-nuclear target reaction one would like to use the same 
definition and for that one needs a picture of 
a nucleus as composed from quasi-free nucleons (Impulse Approximation
- IA \cite{pwia}), like in the Fermi Gas (FG)
model. In the $\sim 1$~GeV energy region typical values of momentum 
transfer are large enough and IA can be used as a reliable approximation. 
However, in inclusive neutrino
measurements there is always a significant fraction of low momentum transfer
{\it apparently CCQE} events and one cannot be sure that they are described in the 
proper way. A remedy is to impose suitable cuts in momentum transfer 
or in $Q^2$, or to use more sophisticated theoretical 
models\footnote{From electron scattering experiments it is known 
that for momentum transfer $q\leq 
350-400$~MeV/c IA based models fail to reproduce the data. In this 
region collective nuclear excitations become important and computational 
techniques like CRPA or RPA should be used \cite{rpa_luis}. Since $q>\omega$ 
($\omega$ is the energy transfer) and $Q^2\equiv q^2-\omega^2>0$, the region of 
the failure of IA is contained in the domain  $Q^2< 0.1$~GeV$^2$. 
In the case of neutrino CCQE interactions evaluated in the IA scheme a region $q\leq 
350-400$~MeV/c corresponds to 
$15\%-20\%$ of the total CCQE cross-section, independently on the 
neutrino energy $E_\nu$ (for $E_\nu<500$~MeV the fraction is even larger) 
\cite{artur_ia}. Experimental groups invented various ad hoc ideas 
to deal with the low $Q^2$ problem. Eg. the MB collaboration introduced a parameter $\kappa$ \cite{kappa} to increase the Pauli blocking effect. }.

Nuclear environment affects CCQE interaction in other ways as well. There 
is a problem of Final State Interactions (FSI): hadrons
arising in a primary interaction must propagate through nucleus 
before they can be 
detected. Thus, for an experimentalist it is natural to speak 
about QE-like events specified by a condition that there are no 
mesons  in
the final state. There is an important difference 
between QE and QE-like events because the latter include  those in which 
a pion produced in the initial interaction was later absorbed 
inside the nucleus. There is also a possibility that for the
CCQE primary interaction nucleon
rescatterings result in pions in the final state. 

\section{Quasielastic axial mass}

A theoretical description of free nucleon target CCQE reaction 
is based on the conserved vector current (CVC) and the partially conserved 
axial current (PCAC) hypotheses. As a result of a simple analysis the
only one unknown quantity is the axial form-factor $G_A(Q^2)$ for 
which one typically 
asssumes the dipole form $F_A(0)(1+\frac{Q^2}{M_A^2})^{-2}$ with 
one free parameter, called the axial mass 
$M_A$. If a deviations from the dipole form of $G_A$ are of a similar size 
as those in the case of electromagnetic form-factors it would be
difficult to detect them and the basic assumptions described above seem to 
be well justified\footnote{
In the early years of neutrino experiments many groups reported also fits 
to the non-dipole axial FF as motivated by quark model vector-dominance: 
$F_A(Q^2)= \frac{F_A(0)}{1+\frac{Q^2}{M_A^2}}\cdot \exp 
\left( - \frac{Q^2[GeV^2]}{1+\frac{Q^2}{4M^2}}\right)$
\cite{non_dipole}. The quality of best fit values for $M_A$ in both 
models was similar. ANL and BNL collaborations considered also
{\it monopole} and {\it tripole} axial FFs but the obtained fits 
were worse then the dipole ones.}. 
Thus, the aim of CCQE experiments is to measure the value 
of $M_A$, the parameter describing free nucleon weak transition matrix element.

Measurements of $M_A$ typically focus on the shape of 
differential cross-section in $Q^2$ which is sensitive 
enough for quite precise evaluations of $M_A$. Investigations of only the 
shape of the $Q^2$ distributions of events do
not rely on the (very limited) knowledge of the neutrino flux. 
The dependence of the total cross-section on $M_A$ 
can also be used as a tool to fix its value. The limiting value of the CCQE 
cross-section $\sigma^{CCQE}_\infty$ as $E_\nu\rightarrow\infty$ can be calculated in the analytical 
way assuming dipole vector and axial form-factors \cite{ankowski}. In the 
exact formula the dependence of $\sigma^{CCQE}_\infty$ on $M_A$ is strictly speaking quadratic but in 
the physically relevant region it is with a good approximation linear. If a value of $M_A$ is increased
from $1.03$ 
to $1.33$~GeV for $E_\nu>1$~GeV the cross-section and (neglecting an impact of detector efficiency 
modifications) the expected number of CCQE events is raised by 
$\sim 30\%$ (for $E_\nu<1$~GeV the increase is smaller). 

In the past several measurements of $M_A$  were done on the 
deuterium target for which serious nuclear physics complications 
are absent. Until a few years ago it seemed that the results 
converge to a value of the order of $1.03$~GeV \cite{bodek_MA}. There is an additional 
argument in favor of a similar value of $M_A$ coming from the weak 
pion-production at low $Q^2$. PCAC based evaluation gives the axial mass value of 
$1.077\pm 0.039$~GeV \cite{MA_PCAC}. On the contrary, all (with an exception 
of the NOMAD experiment) more recent high statistics measurements of 
$M_A$ report much larger values: K2K (oxygen, $Q^2>0.2$~GeV$^2$) 
$\rightarrow 1.2\pm 0.12$ \cite{k2k_oxygen_MA}; K2K (carbon, 
$Q^2>0.2$~GeV$^2$) $\rightarrow 1.14\pm 0.11$\cite{k2k_carbon_MA}; 
MINOS (iron, $Q^2>0$~GeV$^2$) $\rightarrow 1.19\pm 0.17$; MINOS (iron, 
$Q^2>0.3$~GeV$^2$) $\rightarrow 1.26\pm 0.17$\cite{minos_MA}; MiniBooNE 
(carbon, $Q^2>0$~GeV$^2$) $\rightarrow 1.35\pm 0.17$; MiniBooNE (carbon, 
$Q^2>0.25$~GeV$^2$) $\rightarrow 1.27\pm 0.14$ \cite{MB_MA} (for completness: 
NOMAD (carbon, $Q^2>0$~GeV$^2$) $\rightarrow 1.07\pm 0.07$ 
\cite{nomad_MA}).

The difference between MB and NOMAD measurements can hopefully be 
explained by different definitions of the CCQE signal. 
In the case of MB a sample of 2-subevents (Cherenkov light 
from muon and from decay electron) is analyzed and ejected protons 
are not detected. In the case of NOMAD 1-track (muon) and 
2-tracks (muon and proton)
samples of events are analyzed simulateuosly. With a suitable chosen 
value of the formation zone parameter $\tau_0$ \cite{formation_zone} 
values of 
$M_A$ extracted separately from both data samples are approximately 
the same\footnote{When $\tau_0$ is increased FSI effects become more 
suppressed. This makes the predicted number of 1-track events smaller and 
2-track events - larger. Thus, the fitted value of $M_A$ from 2-track 
events becomes smaller and from 1-track events - larger.}

More detail characterization of CCQE scattering was given by the MB 
experiment in the form of double differential cross section in muon 
kinetic energy and opening angle. A subtraction of the background (CCQE-like 
but not CCQE events) was done in the way which was intended to be 
independent on Monte Carlo (NUANCE \cite{nuance}) modelling of 
nuclear effects. 
A correction DATA/MC function was obtained based on the sample of events 
dominated by the pion production and then applied to MC background 
predictions. The shape of the correction function is not well understood 
but it has an 
important impact on the extracted value of $M_A$. The function quantifies 
a lack of precision in describing processes like pion absorption and this 
can have different effect on understanding of QE-like and SPP-like samples 
of events. One can also use the CCQE signal and the background 
together as the measurement of CCQE-like cross section, the
observable which is in the minimal degree dependent on MC assumptions.

NUANCE implementation of the IA is based on the FG model which does not
provide a realistic  
description of the nucleon momenta distrubution. 
This motivated later axial mass fits done within 
more sophisticated nuclear 
models. The authors of \cite{jsz} used spectral function 
\cite{spectral_function} and made a fit to full double differential 
cross section data. The overall normalization error evaluated by MB as 
$10.7$\% was also taken into account. Butkevich\footnote{Target 
nucleon is a solution of the self-consistent Dirac equation in 
the $\sigma -\omega$ theory and ejected nucleon is treated in RDWIA 
(Relativistic Distorted Wave Impulse Approximation). Short Range 
Correlations effects are also taken into account \cite{butkevich_model}} 
made a fit only to the $Q^2$ differential cross section data 
\cite{butkevich}. Both analysis produced similar results: Butkevich 
obtained $1.37\pm 0.05$ and JSZ $1.34\pm 0.06$~GeV (with the low-momentum 
cut $q_{cut}=500$~MeV/c, details in \cite{jsz}).

\section{Multinucleon ejection}

A possible theoretical mechanism which can explain the $M_A$ value discrepancy 
comes from the many-body nuclear model proposed more then 
10 years ago \cite{marteau} and developed later by 
Martini, Ericson, Chanfray and Marteau (MEChM model)
\cite{martini}. 
The idea of the importance of the many-body contribution in neutrino 
interactions is even older and was 
presented by Magda Ericson \cite{ericsson}. The model in \cite{ericsson} 
discusses an appearance of the pion branch, a collective nucleus 
excitation which decays into a pair of nucleons. 

MEChM is the non-relativistic model that includes QE and $\Delta$ production primary interactions, RPA 
corrections and local density effects. Its interesting 
feature is the evaluation (without meson exchange current contribution) of elementary 
2p-2h and 3p-3h excitations which lead to multinucleon ejection. 
This contribution is 
absent in free nucleon neutrino reaction. However, 
the evaluation of the np-nh contribution was 
based on approximate arguments and the  authors of \cite {martini} admitted 
that a detail microscopic computation was missing.
Within the MEChM model the new contribution is shown to be able to account for 
the large CCQE cross-section as measured by the MB collaboration, 
In the case of neutrino-carbon CCQE process (after averaging over the 
MB neutrino beam) the nuclear effects are expected to increase the 
cross-section 
from 7.46 to 9.13 (in the units of $10^{-39}$cm$^2$ per neutron). This 
includes the
cross-section reduction due to RPA effects and a substantial 
increase due to 
np-nh contribution. 
It is interesting that in 
the case of antineutrino-carbon CCQE scattering, because a relative significance of the 
isovector contribution is larger,
RPA and 2p-2h effects cancell each other and the flux averaged cross-section 
remains virtully unchanged (modification from 2.09 to 2.07 
in the units of $10^{-39}$cm$^2$ per proton). 

Shortly before the NuInt11 Workshop the microscopic many-body evaluation of 
multinucleon 
ejection contribution was reported in the paper \cite{nieves_2p2h}.
The computations were done in the theoretical scheme which 
is known to be 
succesfull in describing electron scattering in the kinematical region 
of QE and $\Delta$ excitation peaks together with the 
DIP region between them \cite{gil_nieves_oset}. 
The model of \cite{nieves_2p2h}
includes medium polarization effects and $\pi$ and $\rho$ meson exchange
contributions in the vector-isovector channel. For neutrino scattering predictions of \cite{nieves_2p2h} and \cite{martini} are in good agreement. However, \cite{nieves_2p2h} predicts a substantial increase of the cross section also in the case of antineutrino scattering, contrary to \cite{martini}.

\section{Recent experimental developments}

During NuInt11 Workshop some new (usually preliminary) experimental results were presented. 
MINOS devoted \cite{minos_nuint11} much effort on better 
evaluation of the pion production background. A function of $Q^2$ which  
corrects Monte Carlo (NEUGEN) RES (resonance production region) predictions 
was proposed. The shape of the curve is similar to MiniBooNE's 
DATA/MC correction function but in the case of MB for $Q^2>0.1$~GeV$^2$ 
the correction factor is $>1$. The new MINOS 
best fit value of $M_A$ is $1.16$~GeV and the error was reduced by a factor of $3$.

SciBooNE showed \cite{sciboone_nuint11} partial results of the CCQE analysis. 
Results are given in terms of fits for CCQE cross-section 
DATA/MC multiplicative 
factors $a_j$ ($j$ corresponds to true neutrino energy bins) and also an overall rescaling factor $F_N$. The obtained best fit values in the region $E\in (0.6, 1.6)$~GeV are between $1.00$ and $1.09$ which with $F_N=1.02$ and the value of the axial mass used in the NEUT Monte Carlo generator ($1.2$~GeV)  
will most likely translate to the axial mass value 
$M_A\sim 1.25 - 1.3$~GeV. The problem with SciBooNE measurement is that 
there are some instabilities in the wider range of neutrino energies 
(see Fig. 11.2 in \cite{alcaraz_thesis}). Also a use of only one universal 
background rescaling factor $a_{bcg}$ for three quite different event samples 
seems worrying, particulary because its best fit value is quite large ($1.37$). 

An important antineutrino CCQE measurement was reported by the MiniBooNE 
\cite{miniboone_nuint11}. The DATA/MC average  cross-section ratio 
was measured to be 
$1.21\pm 0.12$ which is a surprising result because in the NUANCE 
carbon CCQE $M_A$ value was $1.35$~GeV. This results (if 
confirmed) may indicate that the treatment of multinucleon ejection 
contribution presented in the paper \cite{nieves_2p2h} is more accurate. 
In the experimental analysis it was very important to evaluate correctly 
a contribution from neutrino contamination in the anti-neutrino flux. 
Three independent measurements indicate that the $\nu_\mu$ flux should be 
scaled down by a factor of $\sim 0.8$ with 
an important impact on the final results.

Preliminary results from the MINERvA experiment \cite{minerva_nuint11} 
with antineutrino beam indicate that events distribution in $Q^2$ is 
slightly below MC predictions (GENIE with $M_A=0.99$~GeV). However, one 
cannot say that there is no large $Q^2$ surplus of CCQE-like events 
because this region is strongly dominated by RES and DIS dynamics.

\section{Solution of the axial mass puzzle?}

In the very recent paper \cite{nieves_2d_fit} the model described in 
\cite{nieves_2p2h} was applied to MB double differential cross section data 
and a fit to the axial mass value was done. Strictly speaking the
model used in the statistical analysis was the one presented in 
\cite{nieves_relativistic} because being a
relativistic one it is more reliable in the whole kinematical region of the
MB experiment. The model does not include FSI diagrams which can introduce
modifications of the size of $\sim 7\%$. In the fitting procedure (taken from
\cite{agostini} and used also in \cite{jsz}) the authors included an 
overall $10.7\%$  normalization error.
The two-parameter fit gave results: $M_A=1.077\pm 0.027$~GeV and
for the normalization scale: $\lambda = 0.917\pm 0.029$. It is 
interesting that
with the low-momentum cut procedure, as proposed in \cite{jsz}, with 
$q_{cut}=400$~MeV the value $M_A=1.007\pm
0.034$~GeV was obtained which is even closer to the historical world average.

\section{Conclusions}

A discussion of CCQE neutrino interaction on nuclear targets becomes quite
complicated because it is necessary to consider the multinucleon 
ejection contribution both on experimental and theoretical levels. 
It seems that
on the theoretical side the situation  becomes clear and there are 
computations
which show that the multinucleus ejection confused with genuine 
CCQE events can lead to large $M_A$ measurements. It is important that as 
the cross-check
the same models are confronted with precise electron scattering data in
kinematical regions similar to that of the MB experiment.
In order to provide an experimental verification of the multinucleon
ejection mechanism
it is necessary to implement the models in MC event 
generators used by experimental groups
and find predictions for kinetic energy of ejected nucleons which can 
be confronted with observables like the vertex activity.




\begin{theacknowledgments}
I would like to thank S.K. Singh and his collaborators for organizing a very 
successful NuInt11 Workshop.
The author was supported by the grants: N N202 368439 and DWM/57/T2K/2007.
\end{theacknowledgments}



\bibliographystyle{aipproc}   


\end{document}